# An Astronomer's View of Climate Change

*by Donald C. Morton*
*Herzberg Astronomy and Astrophysics Programs*
*National Research Council of Canada*
*5071 West Saanich Rd, Victoria BC V9E 2E7, Canada*

## Abstract

This paper describes some of the astronomical effects that could be important for understanding the ice ages, historic climate changes and the recent global temperature increase. These include changes in the Sun's luminosity, periodic changes in the Earth's orbital parameters, the Sun's orbit around our galaxy, the variability of solar activity, and the anticorrelation of cosmic-ray flux with that activity. Finally, recent trends in solar activity and global temperatures are compared with the predictions of climate models.

## 1. Overview

In successive sections, this article will discuss the stability of the Sun's luminosity, how long-term changes ($\gtrsim 10^4$ yr) in the Earth's orbit around the Sun and through the Milky Way Galaxy can affect global temperatures, and how shorter-term changes ($\lesssim 10^4$ yr) in the luminosity or the solar magnetic field through the modulation of galactic cosmic rays can also be important. Later sections compare global temperature measurements with the predictions of the General Circulation Models (GCM) from the Reports of the Intergovernmental Panel on Climate Change (IPCC 2007, 2013).

## 2. The Stability of the Solar Luminosity

The simplest explanation of the ice ages and other climate changes is variations in the Sun's luminosity. The relevant quantity at the Earth is the solar constant or total irradiance—the integrated flux over all wavelengths outside the Earth's atmosphere at the mean distance of one astronomical unit. There has been much effort for nearly two centuries to measure the luminosity, but calibration difficulties have led to large uncertainties, and even satellite instruments recording the whole spectrum since 1978 have had their inconsistencies. With the NASA *Solar Radiation and Climate Experiment* (SORCE), Kopp and Lean (2011) found the value to be 1360.8 ± 0.5 Wm$^{-2}$ during the 2008 solar minimum and similar values for the previous two minima, slightly less than the 1365.4 ± 1.3 Wm$^{-2}$ adopted by most climate models. Note that the models actually use 1365.4/4 = 341.35 Wm$^{-2}$, which is the flux per unit surface area of a spherical Earth.

The new calibration is not expected to have a significant effect on the climate simulations. The mean peak-to-peak change in total irradiance over three solar cycles since 1978 is about 0.12 percent with occasional extremes up to 0.34 percent that could be partially compensated by unseen parts of the Sun. The associated change in global temperature during a cycle is approximately 0.1 °C, so the 0.5 °C warming during the past 35 years cannot be due to an increase in the irradiance.

Were larger variations possible in earlier times? The Sun is a spherical ball of gas in hydrostatic equilibrium with a central temperature of $1.6 \times 10^7$ K, hot enough for nuclear energy generation by converting hydrogen into helium. During an evolutionary time of $8 \times 10^9$ yr on the main sequence, the models of Bahcall *et al*. (2001) show that the luminosity of the Sun will increase gradually by about a factor of two. The lower luminosity in earlier years implies an Earth temperature below the freezing point of water more than $2 \times 10^9$ yr ago, while geological evidence suggests that liquid water has been present for the past $4 \times 10^9$ yr. Extra water vapour or $CO_2$ to trap the heat seems insufficient, so Sackmann & Boothroyd (2003) considered a Sun starting at 1.07 solar masses. Their model produces acceptable temperatures but requires a wind with a thousand times more flux than now to dissipate the extra mass.

On shorter time scales, there is the added complexity of the Sun's magnetic field and the internal mass motions of a dynamo to generate that field as well as turbulence and a convective outer envelope that carries the energy to the surface. Li *et al*. (2003) included these effects in simple models of the solar interior and found the luminosity could vary by the observed 0.1 percent. Much larger fluctuations seem unlikely, but further analysis with more realistic models is needed for a clearer answer. Meanwhile, it is instructive to examine other contributors to the Earth's climate.

## 3. Long-term Astronomical Effects ≳ $10^4$ Years

The Serbian scientist Milankovitch (1941, 1969) investigated the slow variations in the parameters of the Earth's orbit around the Sun as the cause of the ice ages. The semi-major axis and the length of the year are very stable, but changes in other elements due to perturbations by the Moon and planets affect the insolation and its seasonal and geographical distribution. The direction of the Earth's pole precesses with a period of 26 kyr (1000 × year), while the orbital axis precesses over 112 kyr in the opposite direction for a combined period of about 22 kyr. The obliquity of the ecliptic oscillates over 41 kyr, and the orbital eccentricity has periods of about 100 and 413 kyr. Since the $^{18}O$ isotope is enhanced in ocean sediments and shells during colder intervals, it is a useful temperature proxy. When the first three Milankovitch periods were found in seabed cores, there was general acceptance that this orbital forcing could explain the ice ages. However, as seen from Figure 1, there was an unexpected change in the dominant period from 100 kyr to 40 kyr about 800 kyr ago and the 22-kyr signal is less significant, even though the precession term was expected to be the strongest. Furthermore Wunsch (2004) concluded that the Milankovitch terms could explain at most 20 percent of the observed temperature changes.

Alternative explanations have considered the Sun's orbit around our galaxy and where supernovae would be most frequent, because the associated cosmic rays could seed clouds that reflect more sunlight. Shaviv & Veizer (2003) estimated the times when the Sun passed through spiral arms, while Svensmark (2012) looked at close approaches to young clusters. Both studies found coincidences of the strongest cosmic-ray fluxes with colder temperatures indicated by $^{18}$O enhancement, but significant differences in the two results prevent definite conclusions. These interesting ideas deserve further investigation as our knowledge of galactic structure improves.

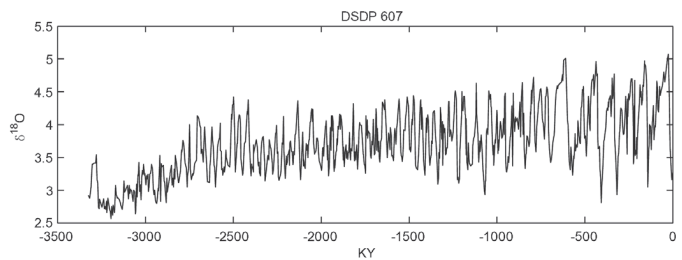

*Figure 1 — Plot from Wunsch (2004) of the $^{18}O$ temperature proxy for the past 3.3 Ma from a subpolar North Atlantic seabed core. Increasing $^{18}O$ implies colder temperatures. The dominant periods are about 100 kyr during the past 800 kyr and 40 kyr for earlier times rather than the expected 22 kyr from the basic Milankovitch theory.*

## 4. Short-term Astronomical Effects ≦10⁴ Years: Radiation and Sunspots

Again in consideration of orbital effects, there is the 2400-year period of the oscillation of the Sun about the barycentre (centre of mass) of the Solar System due to planetary perturbations. Since the barycentre usually is within two solar radii, the fractional variation in insolation is about ±10$^{-4}$ or only 1/10 the mean variation over the solar cycle.

As the Sun rotates with an average period of about 27 days, various active regions come into view from the Earth. The total irradiance can vary by ± 0.34 percent and the ultraviolet (UV) and X-radiation as well as the solar wind and its perturbations of the Earth's magnetosphere can change by larger amounts in a random way. Thus the rotation can add brief spikes to the measurements of these indices.

The principal short-term effect is the ~11 year sunspot cycle discovered by the amateur astronomer Heinrich Schwabe in 1843. Actually, it is a ~22 year period because the new spots appearing at the maximum of a cycle reverse their magnetic polarity each ~11 years. At the peak of solar activity, the 0.1-percent reduction in solar irradiance by the dark spots is more than compensated by 0.2-percent increase from bright faculae or other regions of the solar surface.

In an important paper reviewing the history of sunspot observations, Eddy (1976) described how Rudolf Wolf began a systematic counting soon after Schwabe's discovery and estimated the numbers back to 1700 by searching old records. Later, Gustav Spörer and Edward Maunder extended these historical investigations, demonstrating a real absence of spots and aurorae from about 1645 to 1715. During some decades in this interval, spots were so infrequent that an astronomer could write a paper if he saw one. Galileo and others with early telescopes first reported spots in 1610. Forty years later, they might have missed them altogether. Figure 2 plots the sunspot numbers since 1610 showing what Eddy called the Maunder Minimum and a later one from about 1800 to 1820 now known as the Dalton Minimum.

The Maunder and Dalton minima are especially interesting, because they occurred when global temperatures were unusually cold. During the Little Ice Age from about 1430 to 1850, glaciers advanced in the European Alps, while canals in Holland and the Thames River in London froze during some winters. Grove (2001) concluded from the radiocarbon dating of trees felled by advancing ice around the world that the Little Ice Age was a global phenomenon and began in the 13th century in many places.

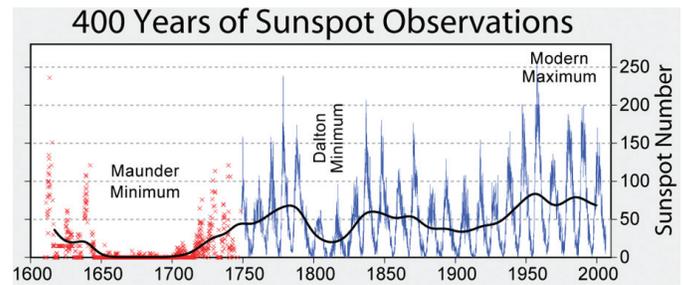

Figure 2 — This plot from the U.S. National Oceanic and Atmospheric Agency shows sunspot numbers since their first observation with telescopes in 1610. Systematic counting began soon after the discovery of the ~11 year cycle in 1843. Later searching of old records provided the earlier numbers.

Further historical investigations have revealed earlier warm intervals now referred to as the Roman (250 BC-AD 400) and Medieval (950-1250) Maxima. The latter matches the Norse settlements in Greenland and the explorations of the east coast of North America at least as far south as L'Anse-aux-Meadows in Newfoundland and a grape-growing region called Vinland. The following colder interval must have contributed to the abandonment of these settlements. All of these dates with warm and cold designations are rather imprecise and have regional differences.

In Switzerland, there is a mountain pass at 2756 m called Schnidejoch between Lenk on the north and Sion on the south side of the Bernese Alps. It was blocked by ice and snow year round and never considered a route joining these towns until an exceptionally warm summer in 2003 opened it to hikers and then archeologists. The carbon dates on 73 artifacts found there ranged from a Neolithic 4800 BC to a Medieval AD 1000 with gaps from 4300 to 3700 BC and 1500 to 200 BC, when ice probably closed the route (Hafner 2012). Temperatures warm enough to melt the ice on Schnidejoch occurred before modern industry and transportation began adding $CO_2$ to our atmosphere. The present recession of glaciers in much of the world confirms we are in another warm period.

It is possible to estimate past temperatures from the growth of tree rings, corals, and ocean sediments, or from deuterium or $^{18}$O isotopes in these deposits because $HD^{16}O$ and $H_2^{18}O$ condense faster than normal $H_2^{16}O$. Mann, Bradley, & Hughes (1998, 1999) used some of these proxies to derive their "hockey stick" plot of Northern Hemisphere temperatures for the past 1000 years. It indicated a gradual decrease with some fluctuations for the first 900 years and then a steep rise beginning in 1900. The absence of the Medieval Maximum and the Little Ice Age raised immediate questions. Canadians Stephen McIntyre of Toronto and Ross McKitrick of the University of Guelph (McIntyre & McKitrick 2003) demonstrated that flawed statistics were used to construct the graph. (See also Essex & McKitrick 2007). Recently Ljungqvist *et al.* (2012) and Christiansen & Ljungqvist (2012) examined a variety of Northern Hemisphere proxies for the past 2000 years. They found that the magnitude of 20th-century warming is within the range of variability over the past 12 centuries, though the present rate of warming has been exceptionally fast. Relative to the interval AD 1880 to 1960, they found a Medieval Warm Period with a peak of +0.6 °C between AD 950 and 1050 and a Little Ice Age cooler by 1.0 °C between AD 1580 and 1720.

One objective measure of solar activity is the radio flux that is Canada's contribution to solar monitoring. Beginning in 1947 in Ottawa, Arthur Covington at the National Research Council (NRC) began a systematic, calibrated measurement of the solar flux at 10.7 cm. This happened to be the band of his war-surplus radar equipment, but it has turned out to be very suitable for studying the Sun. Ken Tapping (2013), at NRC's Dominion Radio Astrophysical Observatory near Penticton, B.C., has continued this monitoring with support from the Canadian Space Agency and participation by Natural Resources Canada. Figure 3 shows the variation in this radio flux over six 11-year cycles. There is no direct effect on the Earth, but the radio emission is a good measure of the ultraviolet flux around 120 nm and it correlates well with the sunspot number, giving a consistent measure of solar variability since 1947 and providing a definite measure of activity when there are no spots. At the 2009 minimum, the monthly average 10.7-cm flux was 4 percent lower than the previous minima. Note also the extra breadth of the last minimum and the weakness of the maximum just past. The sunspot count shows a similar pattern, possibly implying that the Sun could be heading for another Dalton or even a Maunder Minimum.

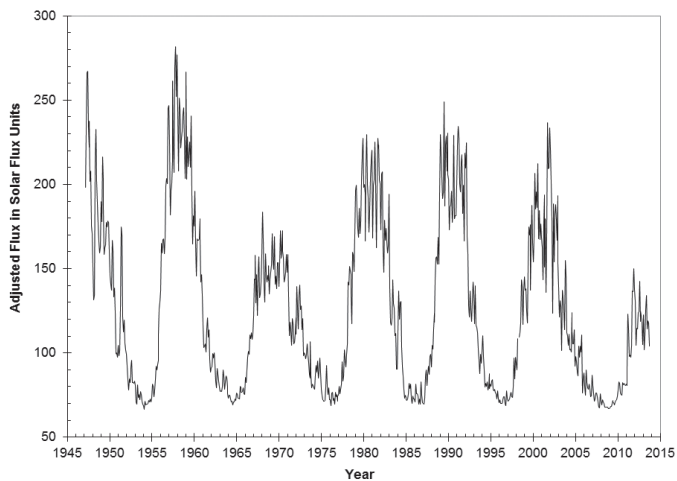

*Figure 3 — Monthly averages of the 10.7-cm solar radio flux measured by the National Research Council of Canada and adjusted to the mean Earth-Sun distance. A solar flux unit = $10^4$ Jansky = $10^{-22}$ $Wm^{-2}$ $Hz^{-1}$. The maximum just past is unusually weak and the preceding minimum exceptionally broad. Graph courtesy of Dr. Ken Tapping of NRC.*

## 5. Short-term Astronomical Effects ≲$10^4$ years: Cosmic Rays

Protons, electrons, ionized atoms, and the attached magnetic field in the solar wind stream out from the Sun at velocities of a few hundred km/s to form the heliosphere. Beyond the orbits of Neptune and Pluto, this wind establishes a pressure balance with the interstellar medium in a region called the heliopause. Any direct effect of the wind on the Earth's climate remains uncertain, but the Sun's magnetic field, with its modulation of cosmic rays reaching the Earth, could be an important astronomical variable influencing climate. This field, which changes polarity every ~11 years and is distorted by the Earth's field and the Earth's motion through the interstellar medium, provides a variable shield against galactic cosmic rays, which can cause mutations in living cells, and and can seed nuclei for clouds. The incident protons (≈87 percent), alpha particles (≈12 percent), and heavier nuclei (≈1 percent) (Sherer *et al.* 2006) collide with atoms in our upper atmosphere to produce a large variety of secondaries including rare isotopes such as $^2D$, $^{13}C$, and $^{18}O$, which can be proxies for temperature. The Sun also emits cosmic rays, but they are weaker and are partly shielded by the Earth's magnetic field.

| Reaction | Half life (yr) | Product |
|---|---|---|
| n + $^{14}N$ → p + $^{14}C$ | 5730 | $^{12}C$ |
| Spallation of 14N, 16O → $^{10}Be$ | 1.51×$10^6$ | $^8Be$ |
| Spallation of $^{40}Ar$ → $^{36}Cl$ | 3.08×$10^5$ | $^{35}Cl$ |

*Table 1 — The Cosmogenic Nuclides*

Especially useful among secondaries are the radioactive nuclei $^{10}Be$ and $^{14}C$, formed by the processes summarized in Table 1. The $^{14}C$ combines with atmospheric $^{16}O_2$ to form $^{14}C^{16}O_2$ that circulates for about 5 years before photosynthesis deposits it as $^{14}C$ in annual tree rings. The $^{10}Be$ isotope attaches to aerosols (any liquid or solid particle suspended in air) that precipitate after about a year. In polar regions, the snow becomes compressed in annual layers of ice. Thus we have two useful histories of the cosmic-ray flux reaching the Earth and the associated solar activity. Solanki *et al.* (2004) found from 11,400 years of $^{14}C$ data that the Sun has been exceptionally active over an unusually long duration from 1940 to 2000. The previous comparable activity was more than 8000 years ago.

Figure 4 from Frölich & Lean (2004) shows how these records correlate with the sunspot numbers—high flux during the Dalton Minimum and even higher during the Maunder Minimum. The $^{14}C$ variations are well known as the de Vries Effect in the dating of organic samples, because it is an important correction to the assumption that the flux is constant in time. The $^{14}C$ and $^{10}Be$ data also show earlier fluctuations with peaks on either side of the Medieval

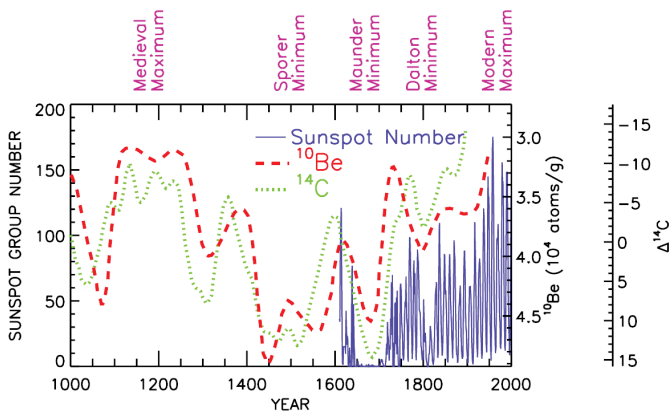

*Figure 4 — The $^{10}$Be and $^{14}$C indices (increasing downward) show the variation in the strength of galactic cosmic rays reaching the Earth and their anticorrelation with sunspot number. The graph is reproduced from Frölich & Lean (2004).*

Maximum and a broad peak from 1420 to 1570 named after Spörer. Except for an interval around 1600, the cosmic-ray flux was particularly strong from 1400 to 1700, coinciding with much of the Little Ice Age.

A remarkable correlation of ocean temperature and cosmic rays follows from the analysis of cores of layered sediments in the North Atlantic by Bond *et al.* (2001). Icebergs carrying debris from glaciers in Canada and Greenland drifted south until they melted at latitudes dependent on the water temperature, leaving signals in the sediments. Figure 5 shows the correlation of this temperature proxy with the $^{14}$C and $^{10}$B fluxes. Meanwhile, the concentration of atmospheric carbon dioxide changed by less than 8 percent.

How can cosmic rays affect global temperatures? Many years ago, Ney (1959) suggested that cosmic rays could provide condensation nuclei to help form clouds. This was a direct extension of the principle of the cloud chamber invented by C.T.R. Wilson, a meteorologist interested in cloud formation, to track ionizing particles. Svensmark & Calder (2007) have described the development of this idea and experiments to test it. When the solar magnetic field is weak, more cosmic rays reach the Earth. Ionization by secondary galactic cosmic rays with 10-20 GeV energies assists in the formation of aerosols of sulphuric acid and water with radii of 2-3 nm. Some of these are hypothesized to grow into cloud condensation nuclei with radii greater than 50 nm that enhance the formation of low-altitude clouds. These clouds reflect sunlight, letting the Earth cool. Laboratory tests at CERN with a pion beam support the first step of this process (Kirkby *et al.* 2011), and experiments in Copenhagen with gamma-ray and natural ionizing sources support the second step (Svensmark, Enghoff & Pedersen 2013). Research is continuing to determine whether these effects are sufficient to influence climate. During the present solar maximum the cosmic-ray fluxes at neutron monitors in Greenland and Finland have been exceptionally strong.

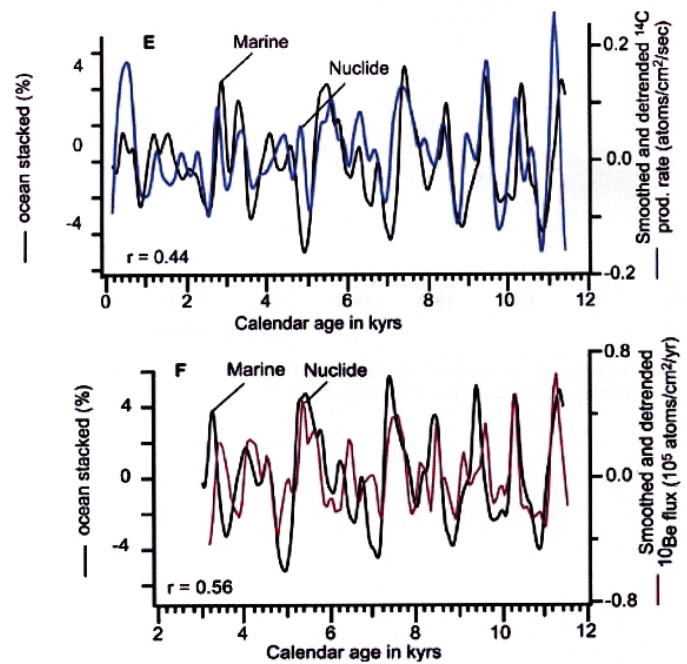

*Figure 5 — Cores of layered ocean sediments indicate the latitude where ice floes melted leaving debris from northern glaciers. In this figure from Bond et al. (2001), the latitude of melting increases towards the top, so temperature increases downwards. Thus stronger cosmic-ray fluxes measured by the $^{14}$C and $^{10}$Be isotopes correlate with cooler North Atlantic temperatures.*

Of course, it is possible that the past cold intervals resulted entirely from a reduced solar irradiance either at visible or UV wavelengths, so that a stronger flux of cosmic rays is just an indicator of an inactive Sun with lower total luminosity. However, the amplitudes of the fluctuations required for the ice ages seem greater than expected for the Sun, so alternative explanations such as the cloud-seeding hypothesis deserve further investigation.

Another possibility is a variation in the ultraviolet solar flux. From minimum to maximum in recent solar cycles, the radiation near 200 nm that produces ozone ($O_3$) from oxygen in the stratosphere increased by as much as 6 percent, and between 240 and 320 nm in the $O_3$ absorption bands, by up to 4 percent, while the total flux increased by only 0.12 percent. Gray *et al.* (2010) describe current research into mechanisms for coupling this heating into the lower atmosphere. Thus temperatures could go up and down with solar activity with minimal change in the total irradiance.

## 6. Current Global Temperature Trends and the Atmospheric Models

Since the recent pattern of solar activity suggests a cooler Earth, it is interesting to examine what has happened to the global temperature. Several institutions have produced graphs of changes in global annual temperatures by calculating the anomaly, the difference from a long-term mean at each location and averaging over continents, oceans, and seasons.

Canadians Christopher Essex of Western University and Ross McKitrick, already mentioned, have raised serious questions about both the physical significance of a mean temperature for a nonequilibrium Earth and the variability of the results, depending on the choice of the statistical procedures. Essex & McKitrick (2007) also discussed the inadequacies of the modelling process, but this paper will continue to refer to these temperatures and models, because they are central to the IPCC reports.

Figure 6 is one such plot of the mean temperature. The recent rise, beginning about 1977, continued until 1998 with no significant increase since. Already in 2009, the change in slope was a concern. At that time, Knight *et al*. (2009) asked the rhetorical question "Do global temperature trends over the last decade falsify climate predictions?" Their response was "Near-zero and even negative trends are common for intervals of a decade or less in the simulations, due to the model's internal climate variability. The simulations rule out (at the 95% level) zero trends for intervals of 15 yr or more, suggesting that an observed absence of warming of this duration is needed to create a discrepancy with the expected present-day warming rate."

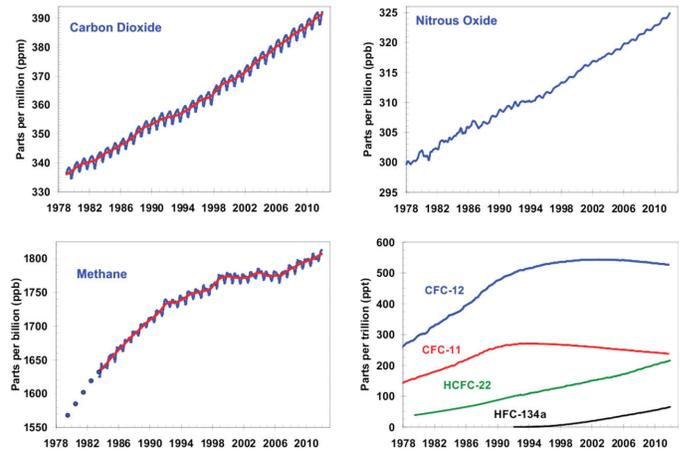

Figure 7 — Measured concentrations of the greenhouse gases $CO_2$, $CH_4$, $N_2O$ and the four most abundant chlorofluorocarbons, all from the global air sampling network of the U.S. National Oceanic and Atmospheric Agency. Increasing uptake of $CO_2$ during Northern Hemisphere spring and summer produces the annual oscillations. However, the worldwide slowing of economic activity following the 2008 recession had no discernable effect.

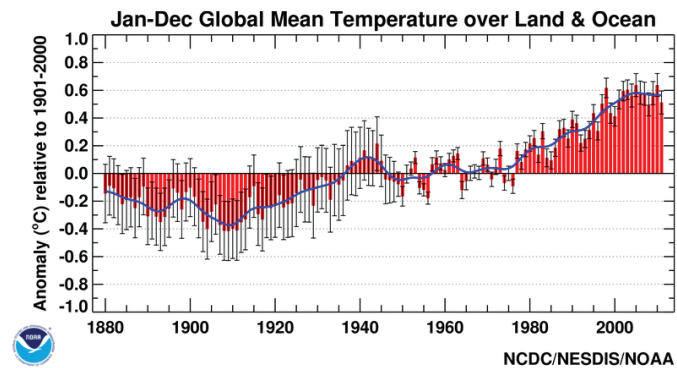

Figure 6 — Recent history of the global mean temperature anomaly compiled by the U.S. National Oceanic and Atmospheric Administration. Note the absence of any rise since 1998.

Now we have 16 years with no indication of increasing global temperature, while the atmospheric concentrations of $CO_2$ and other greenhouse gases shown in Figure 7 continue to rise. Figure 8 compares temperature observations with the predictions of an average of 42 models. By the 15-year criterion, the IPCC models have failed the test of predicting the temperature. Some climate scientists now are saying that 15 years is too short a time for a test, but if that is the case, the rise over 21 years also could be an aberration. The critical test of any scientific theory or model is whether it makes correct predictions; simply fitting a model to existing data does not validate it for future trends. Unfortunately, a lot of public policy is based on those unsuccessful predictions.

Although other observations such as sea levels indicate continued warming, it is important to understand the surface-temperature hiatus, because environmental advocates and government policies emphasize it and have proposed a goal of limiting the rise to 2 °C. The extent of polar sea ice is less useful: the Arctic is notoriously variable; a previous major retreat occurred in the 1930s; and most of Antarctica has continued to cool.

The IPCC climate models are among the most sophisticated of all computer simulations. Nevertheless, something seems to be missing. The usual explanation of climate begins with a solar flux of 340 Wm$^{-2}$ (on the surface-area scale noted in Section 2) incident on an Earth with no atmosphere that produces a mean temperature of -18 °C. Then, with an atmosphere, the water vapour along with small additional contributions from the so-called greenhouse gases (GHG)—carbon dioxide ($CO_2$), methane ($CH_4$), nitrous oxide ($N_2O$), ozone ($O_3$), and chlorofluorocarbons —absorb about 240 Wm$^{-2}$, heating the Earth to a mean of about +15 °C and reradiating thermally in proportion to the fourth power of the temperature from near unit optical depth into the atmosphere. The remaining 100 Wm$^{-2}$ is reflected from clouds and the surface. Such backwarming is well known to astronomers, who call it line blanketing in their calculations of stellar atmospheres (Mihalas & Morton 1965). This heating by gaseous absorption is well understood, so any explanation of the present constancy of global temperatures must include compensation for the expected continuing warming from increasing concentrations of the greenhouse gases.

Note, however, that absorption is not the cause of heating inside a glass house. Instead the air becomes hot from contact with the heated ground or plants and remains so because it does not circulate with external air that has not been heated. Greenhouse operators often increase the $CO_2$ concentration, because the plants will grow better with extra food!

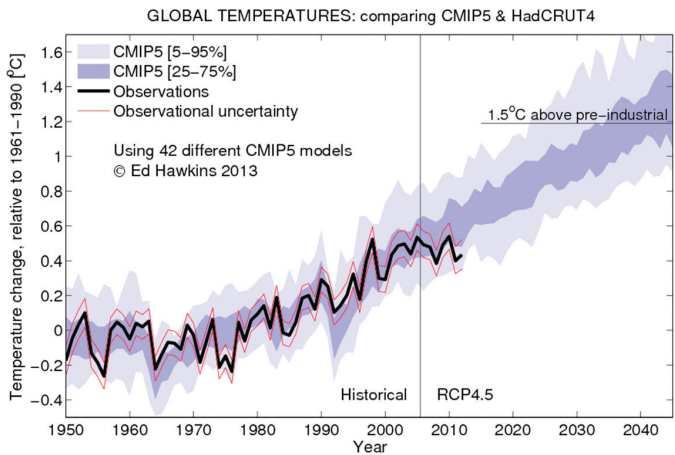

*Figure 8 — Comparison of the average of 42 models from the Coupled-Model Intercomparison Project (CMIP) with the HadCRUT3 mean of global temperatures compiled by the Hadley Centre of the UK Met Office. Graph courtesy of Dr. Ed Hawkins (2013), Dept. of Meteorology, University of Reading, U.K.*

Since the Industrial Revolution, global temperatures have been rising along with increased concentrations of $CO_2$ and the other greenhouse gases. Although coincident behaviour is not proof of a physical connection, the IPCC models made the plausible assumption that $CO_2$ has been the primary cause of the warming. Reflections by clouds and other aerosols approximately cancelled the effects of the other gases. However, the backwarming by $CO_2$ produced only about half the temperature rise, so the models needed to include amplification by a positive feedback caused by hotter air holding more water vapour, which absorbs more radiation. The computer simulations approximated this feedback and many other effects through adjustable parameters to match the observed temperature rise and produce a range of future scenarios. Essentially the feedback has been calibrated by the past rise in temperature; if it is no longer rising, the effect of more $CO_2$ will be less serious.

## 7. The 2013 Report to the IPCC

The latest report IPCC (2013), which was released on 2013 September 28 after the completion of this paper, changes little. The writers recognize the 15-year absence of warming and speculate it is part of natural variability in climate and is biased by a warm El-Niño event in 1998. (Until this explanation was required, the 1998 peak was just another example of warming by $CO_2$.) The report also suggests the deviation from the simulations could be due to heat being absorbed in the oceans or that some models overestimate the effect of greenhouse gases and notes that models can have decade-long intervals of near-constant temperature. The changes in slope for the curves for $CH_4$ and two of the chlorofluorocarbons in Figure 7 also could be contributing, as proposed by Estrada, Perron, & Martinez-Lopez (2013). The IPCC report broadens the range of the predicted temperature increase to 1.5 to 4.5 °C from the previous 2.0 to 4.5 °C for a doubling of the $CO_2$ concentration, thus allowing for a little less warming while retaining the alarming upper limit. Otherwise, the report ignores the change in slope of the temperature curve and a possible clue to some overlooked physics of climate change. If there is a heating bias in some models, why did the upper limit on temperature remain the same? Regrettably, there is no recognition of the significant decrease in solar activity during the last decade.

Also in September, Canadian scientists Fyfe, Gillet, & Zwiers (2013) at the University of Victoria published an important paper confirming the discrepancy between the models and temperatures. The authors averaged only model temperatures at locations coincident with observations and found a predicted rise of 0.21 ± 0.03 °C per decade since 1998 compared with an observed 0.05 ± 0.08 °C. As these authors have noted, there is much work ahead to identify the causes of the discrepancy, construct new models, and see how well they predict temperatures in coming decades.

## 8. Summary and Viewpoint

In summary, at this stage of our understanding, the most important contributors to climate change in order of decreasing time scales are:

a) the gradual evolutionary brightening of the Sun over almost $5\times10^9$ yr,

b) changes in the total solar irradiance, with limitations to be determined from satellite measurements and further solar modelling,

c) a change in the ultraviolet irradiance,

d) changes in cloud formation due to galactic cosmic rays, both in the long term, depending on the proximity of supernovae in the Sun's galactic orbit, and on shorter time scales, depending on the strength of the Sun's magnetic shield, and

e) changes in the concentrations of absorbing gases, both water vapour and the greenhouse gases.

This astronomer's view includes the following thoughts:

1) The present climate models have predicted rising global temperatures that actually have been constant since 1998.

2) Like the astrophysical models described above, climate models can be helpful to test hypotheses and understand physical processes. However, more development is needed before the models are reliable enough for future planning. Without a clear understanding of the temperature plateau, it is impossible to decide how much reduction in anthropogenic greenhouse gases is needed to limit the temperature rise to 2 °C.

3) As we increase the concentration of $CO_2$ and other anthropogenic gases, temperatures could rise again, or they could decrease if the weak solar activity leads to another Maunder Minimum.

4) We should avoid claiming that climate science is settled or that we know how to control the climate.

5) We must beware of science by consensus. There once was general agreement that the ether was necessary for the propagation of light in a vacuum. Science progresses by skepticism and the comparison of theory with experiments and observations.

## Acknowledgements

The author thanks Dr. Jan Veizer of the University of Ottawa and Dr. Richard Goody of Harvard University for helpful comments. ✸